\documentclass[journal,a4paper,onecolumn]{IEEEtran}

\usepackage{tikz}
\usepackage{epsfig}
\usepackage{epstopdf}
\usepackage{siunitx}
\usepackage{pbox}
\usepackage{makecell}
\usepackage{multirow}
\usepackage{amssymb}
\usepackage{amsmath}
\usepackage[normalem]{ulem}
\usepackage{hyperref}

\usepackage{blindtext}
\usepackage{etoolbox}
\usepackage{float}
\usepackage{subfig}

\makeatletter

\docsvlist{\@oddhead,\@evenhead,\ps@headings,\ps@IEEEtitlepagestyle,\ps@IEEEpeerreviewcoverpagestyle}
\makeatother

\begin{document}

\title{A small-signal description of black phosphorus transistor technologies for high-frequency applications}

\author{Leslie Valdez-Sandoval, Eloy Ram\'i­rez-Garc\'ia, Saungeun Park, Deji Akinwande, David Jim\'enez, Anibal Pacheco-Sanchez 

\vspace{-6mm}

\thanks{Manuscript received March 24, 2021; revised April 27, 2021 and May 18, 2021; accepted May 30, 2021. This work has received funding from the European Union's Horizon 2020 research and innovation programme under grant agreements No GrapheneCore2 785219 and No GrapheneCore3 881603, from Ministerio de Ciencia, Innovaci\'on y Universidades under grant agreement RTI2018-097876-B-C21(MCIU/AEI/FEDER, UE). This  article  has been partially  funded  by  the  European  Regional  Development  Funds  (ERDF)  allocated  to  the  Programa Operatiu FEDER de Catalunya 2014-2020, with the support of the Secretaria d'Universitats i Recerca of the Departament d'Empresa i Coneixement of the Generalitat de Catalunya for emerging technology clusters to  carry  out  valorization  and  transfer  of  research  results.  Reference  of  the  GraphCAT  project:  001-P-001702. This article has also received funding from Instituto Polit\'ecnico Nacional under the contract no. SIP/20210167. \newline \indent L. Valdez and E. Ram\'irez are with Instituto Polit\'ecnico Nacional, UPALM, Edif. Z-4 3er Piso, Cd. de M\'exico, 07738, M\'exico, e-mail: lvaldezs1001@alumno.ipn.mx \newline \indent S. Park was with The University of Texas at Austin, Austin, Texas 78758, United States. He is now with Texas Instruments, Dallas, TX. USA. \newline \indent D. Akinwande is with the Microelectronics Research Center, Department of Electrical and Computer Engineering, The University of Texas at Austin, Austin, Texas 78758, United States. \newline \indent D. Jim\'enez and A. Pacheco are with the Departament d'Enginyeria Electr\`{o}nica, Universitat Aut\`{o}noma de Barcelona, Bellaterra 08193, Spain, e-mail: AnibalUriel.Pacheco@uab.cat}
}

\maketitle
\makeatletter
\def\ps@IEEEtitlepagestyle{
  \def\@oddfoot{\mycopyrightnotice}
  \def\@evenfoot{}
}
\def\mycopyrightnotice{
  {\footnotesize
  \begin{minipage}{\textwidth}
  \centering
This work has been accepted to the IEEE for publication. 
© 2021 IEEE. Personal use of this material is permitted. Permission from IEEE must be obtained for all other uses, in any current or future media, including reprinting/republishing this material for advertising or promotional purposes, creating new collective works, for resale or redistribution to servers or lists, or reuse of any copyrighted component of this work in other works. \\DOI: \href{https://ieeexplore.ieee.org/document/9446518}{10.1109/LMWC.2021.3086047}
  \end{minipage}
  }
}
\begin{abstract}
\boldmath
This work presents a small-signal high-frequency (HF) equivalent circuit (EC) to model AC performances of black-phosphorous field-effect transistors (BPFETs). The proposed EC is able to describe correctly both the experimental HF intrinsic and extrinsic figures of merit, as well as S-parameters, from different BPFET technologies. Single- and double-stage radio frequency gain amplifiers, are designed at \SI{2.4}{\giga\hertz} using the experimentally-calibrated small-signal BPFET EC. Results show high-gain high-selective BPFET-based amplifiers.
\end{abstract}


\IEEEpeerreviewmaketitle
\vspace{-5mm}
\section{Introduction}
\label{ch:intro}

During recent years, proof-of-concept low-power HF BPFETs have been demonstrated with both extrinsic cutoff frequency $f_{\rm t,e}$ and extrinsic maximum oscillation frequency $f_{\rm max,e}$ within the range of some tens of \SI{}{\giga\hertz} \cite{WanWan14}-\cite{LiXio20}. One of the technology issues to overcome in BPFETs in order to boost their dynamic capabilities is the high contact resistance associated to Schottky-like barriers at the metal contact and 2D channel interface \cite{LiuNea14}-\cite{YanCha17}. Despite their optimistic HF performance projection \cite{YinAlM17}-\cite{ValRam20} as well as their experimentally HF performance proven potential, there is a lack of RF BPFET-based circuits \cite{LiTia18}, \cite{ChoYog16}. This can be alleviated by providing reliable device HF models. In the literature, the HF figures-of-merit (FoM) of BPFETs \cite{WanWan14}-\cite{LiXio20} have not been described so far by any model whereas a unique small-signal proposal has been compared only with numerical device simulation results \cite{YinAlM17}. In this work, a small-signal equivalent circuit (EC) of BPFETs has been proposed and experimentally validated for the first time across different fabricated technologies and the calibrated models have been used to design RF circuits.

\section{Small-signal description of BPFETs} \label{ch:s_II}

\begin{figure}[!htb] 
\centering
\includegraphics[width=0.47\textwidth]{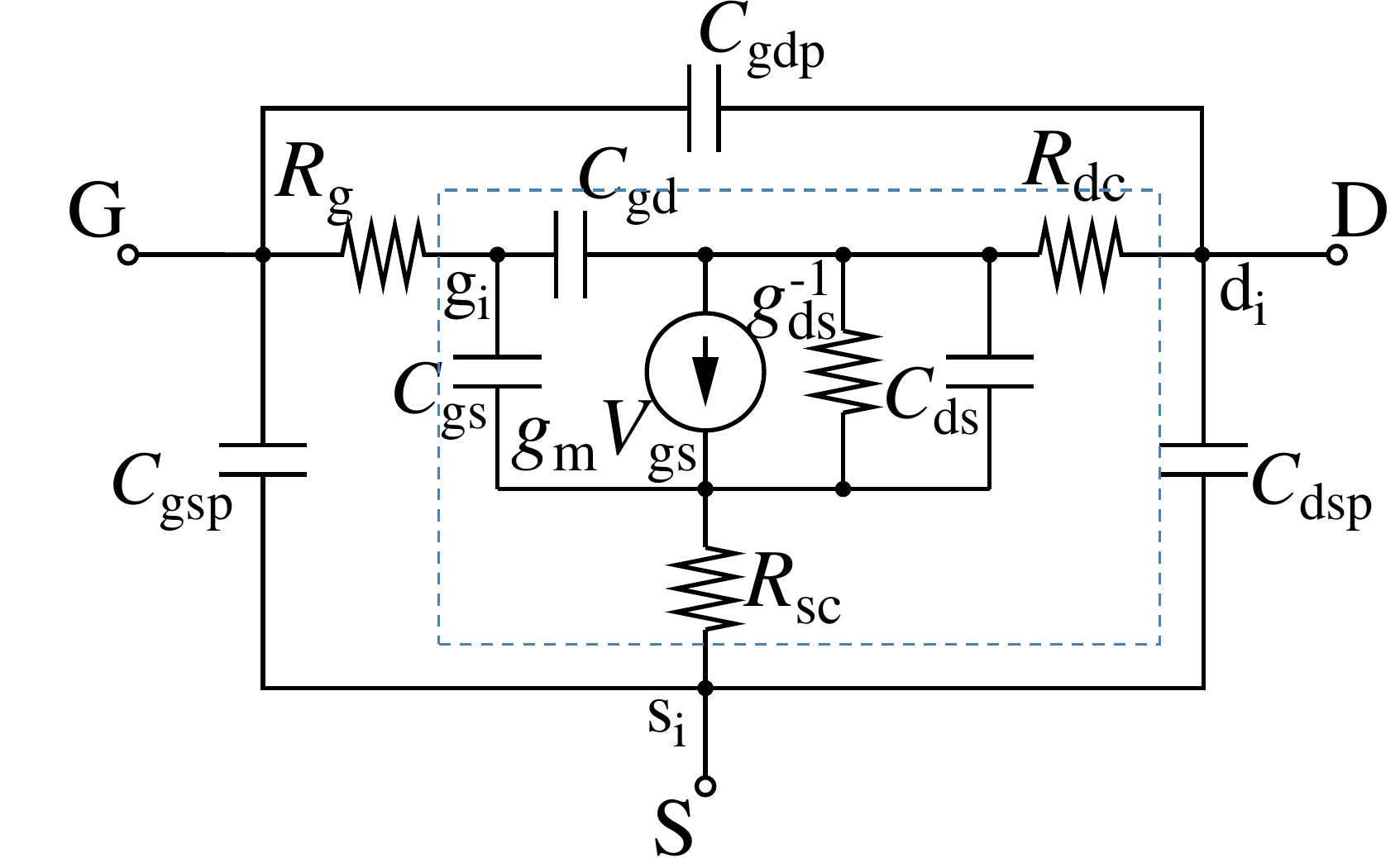}
\caption{\small{Small-signal equivalent circuit model of black phosphorus FETs. The intrinsic transistor is within the dashed box}}
\label{fig:ckt}
\end{figure}

A quasi-static (QS) small-signal EC model of BPFETs, without the parasitic contributions of the access pads, is shown in Fig. \ref{fig:ckt}. 

The intrinsic elements in this approach, which are located between the nodes $\rm g_{\rm i}$, $\rm s_{\rm i}$ and $\rm d_{\rm i}$, represent all internal phenomena. Drain and source contact resistances, $R_{\rm dc}$ and $R_{\rm sc}$, respectively, have been included in the intrinsic device model since they can not be removed with standard de-embedding techniques: they embrace the contribution of bias-dependent potential barriers at the interfaces of the BP channel and the metallic contacts \cite{LiuNea14}-\cite{YanCha17}. Their contribution can be mathematically removed however, if experimental S-parameters are available \cite{RamPac20}, \cite{PacRam21}. Intrinsic electrostatics couplings are represented by the gate-to-source/gate-to-drain/drain-to-source capacitance $C_{\rm gs/gd/ds}$. $g_{\rm ds}$ is the intrinsic output conductance, $g_{\rm m}$ the intrinsic transconductance and $V_{\rm{gs}}$ the intrinsic gate-to-source voltage. The extrinsic model consists of the gate resistance $R_{\rm g}$ and the parasitic gate-to-source/gate-to-drain/drain-to-source capacitance $C_{\rm gsp/gdp/dsp}$. Pad-related parasitics not included in the model can be removed experimentally by on-wafer de-embedding methods \cite{TieHav05}. Gate fingers-related inductance is negligible for emerging transistor technologies.

The model proposed here describes the performance of fabricated RF BPFETs with gate lengths of \SI{250}{\nano\meter} and \SI{500}{\nano\meter} from a technology 1 ($T_{\rm 1}$) \cite{ZhuPar16} and of \SI{300}{\nano\meter} \cite{WanWan14} and \SI{400}{\nano\meter} \cite{LiTia18} from technologies 2 ($T_{\rm 2}$) and 3 ($T_{\rm 3}$), respectively. Devices layout and fabrication details have been presented elsewhere \cite{WanWan14}-\cite{LiTia18}. Small-signal parameters have been obtained at the operating bias point at which the pad de-embedded HF FoMs have been measured in \cite{WanWan14}-\cite{LiTia18}. $R_{\rm dc}$ and $R_{\rm sc}$ have been extracted with a $Y$-function method \cite{ValRam20} considering an underlying transport equation without simplifications \cite{PacJim20}. $R_{\rm g}$ values have been calculated using the approach presented in \cite{VanGee94}. Model parameters for the $T_{\rm 1}$ \SI{500}{\nano\meter}-long device \cite{LiTia18} have been extracted using a proper detachment of $R_{\rm dc/sc}$ \cite{RamPac20}, \cite{PacRam21} enabled by available experimental S-parameters. For the other devices, S-parameters data are not available and hence, the parameters extraction is as follows. $g_{\rm m}$ and $g_{\rm ds}$ have been obtained from the experimental curves provided in the corresponding works \cite{WanWan14}-\cite{LiTia18}. Intrinsic and extrinsic capacitances are obtained by a TCAD-based least-square error-function algorithm \cite{Laux85} towards a correct model description of the experimental HF FoMs reported in \cite{WanWan14}-\cite{LiTia18}. Model parameters for the studied BPFETs \cite{WanWan14}-\cite{LiTia18} are listed in Table \ref{tab:param}. Low values of intrinsic capacitances obtained here are within the same range of the ones obtained for BPFETs have been with another more intricate and experimentally-validated approach \cite{YarHar20}.

\begin{table} [!htb] 
\begin{center}
\caption{\small{Parameter values of the small-signal EC shown in Fig. \ref{fig:ckt} for BPFET technologies with different gate lengths.}}
		\scalebox{0.87}{
\begin{tabular}{c||c|c||c}

	Parameter & \makecell{$T_{\rm 1}$ \cite{ZhuPar16}\\\SI{250}{\nano\meter} (\SI{500}{\nano\meter})\\$V_{\rm GS}=\SI{-1.5}{\volt}$ ($\SI{-3.7}{\volt})$,\\$V_{\rm DS}=\SI{-0.5}{\volt}$ ($\SI{-1.8}{\volt})$} & \makecell{$T_{\rm 2}$ \cite{WanWan14}\\\SI{300}{\nano\meter}\\$V_{\rm GS}=\SI{-1.8}{\volt}$,\\$V_{\rm DS}=\SI{-0.5}{\volt}$}  & \makecell{$T_{\rm 3}$ \cite{LiTia18}\\\SI{400}{\nano\meter}\\$V_{\rm GS}=\SI{-1.5}{\volt}$,\\$V_{\rm DS}=\SI{-3}{\volt}$}  \\ \hline 
				
	\multicolumn{4}{c}{intrinsic part} \\ \hline
				
	$g_{\rm{m}} (\SI{}{\micro\siemens})$ & \SI{41.3}{} (\SI{2500}{}) & \SI{37.2}{} & \SI{34.4}{}   \\ 
	$g_{\rm{ds}} (\SI{}{\micro\siemens})$ & \SI{5.9}{} (\SI{486}{}) & \SI{2.8}{} & \SI{0.1}{}   \\ 
	$C_{\rm{gd}} (\SI{}{\femto\farad})$ & \SI{1.9}{} (\SI{2.6}{}) & \SI{1.6}{} & \SI{1.1}{}  \\
	$C_{\rm{gs}} (\SI{}{\femto\farad})$ & \SI{0.8}{} (\SI{0.7}{}) & \SI{0.7}{} & \SI{0.6}{}  \\
	$C_{\rm{ds}} (\SI{}{\femto\farad})$ & \SI{1.8}{} (\SI{5.8}{}) & \SI{1.1}{} & \SI{0.3}{}  \\ 
	$R_{\rm{sc}} (\SI{}{\kilo\ohm})$ & \SI{30.5}{} (\SI{0.566}{}) & \SI{1}{} & \SI{2.9}{}   \\
	$R_{\rm{dc}} (\SI{}{\kilo\ohm})$ & \SI{30.5}{} (\SI{0.566}{}) & \SI{1}{} & \SI{2.9}{} \\ \hline
				
	\multicolumn{4}{c}{extrinsic part} \\ \hline
				
	$C_{\rm{gdp}} (\SI{}{\femto\farad})$ & \SI{0.7}{} (\SI{3.6}{}) & \SI{0.6}{} & \SI{0.3}{}  \\ 
	$C_{\rm{gsp}} (\SI{}{\femto\farad})$ & \SI{0.5}{} (\SI{13.4}{}) & \SI{0.2}{} & \SI{0.1}{}  \\ 
	$C_{\rm{dsp}} (\SI{}{\femto\farad})$ & \SI{0.06}{} (\SI{1.5}{}) & \SI{0.05}{} & \SI{0.02}{}  \\ 
	$R_{\rm{g}}   (\SI{}{\ohm})$ & \SI{29.3}{} (\SI{29.3}{})& \SI{12.2}{} & \SI{10}{} \\ 
				
	\end{tabular} \label{tab:param}
	}
	\end{center}
\end{table} 

\vspace{-6mm}
\begin{figure}[!htb] 
	\centering
	\hspace{-0.15cm}
	\subfloat{\includegraphics[height=0.415\textwidth]{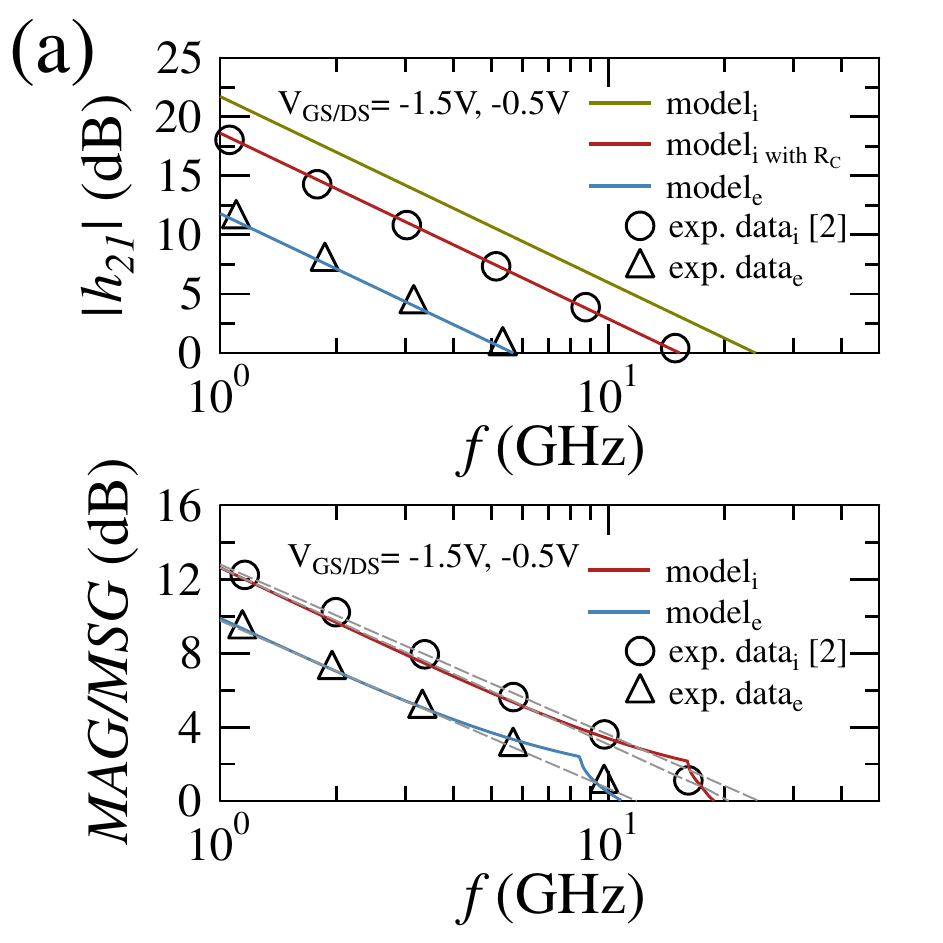}}
	\hspace{5mm}
	\subfloat{\includegraphics[height=0.415\textwidth]{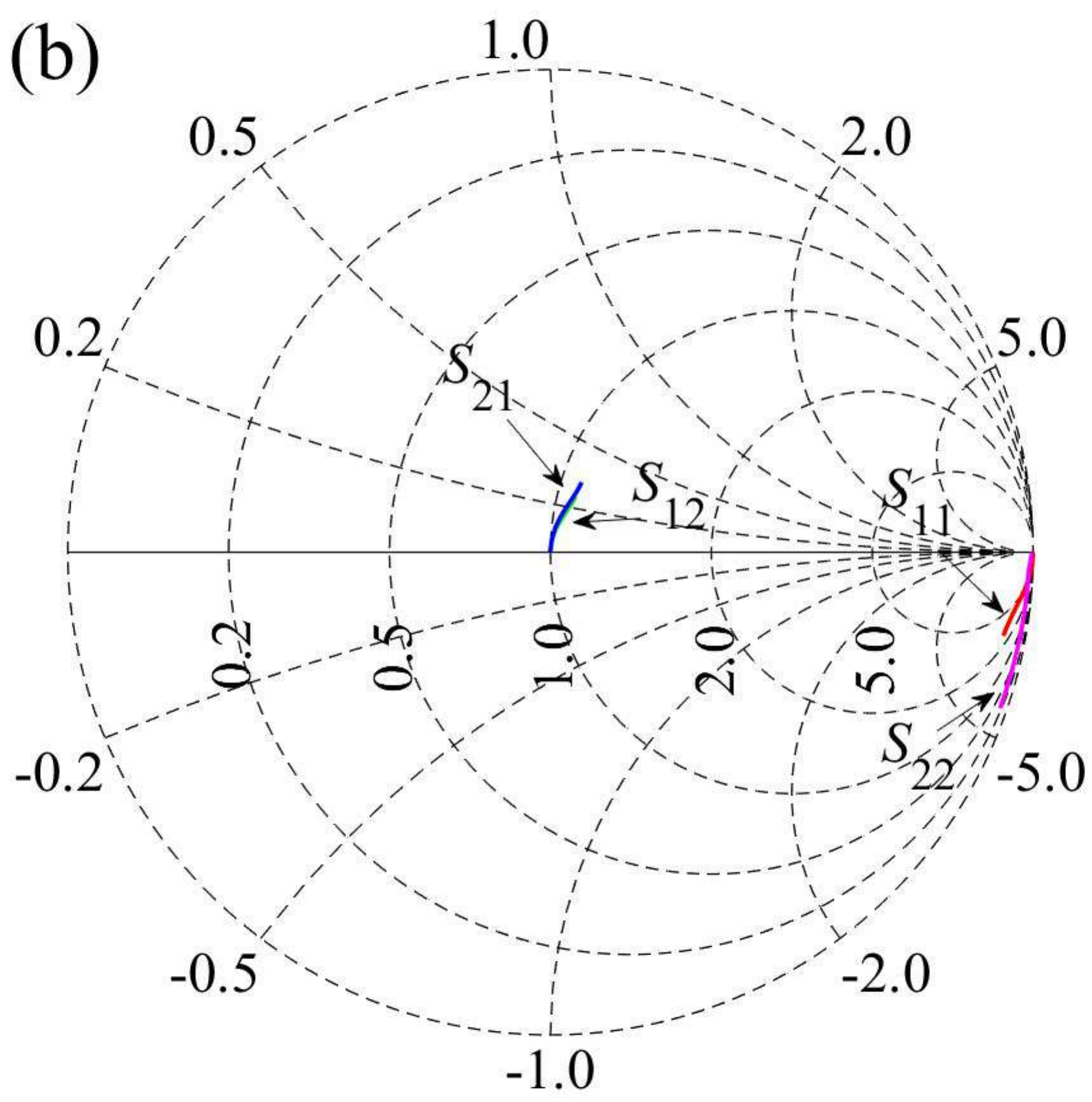}}
	
	\subfloat{\includegraphics[height=0.415\textwidth]{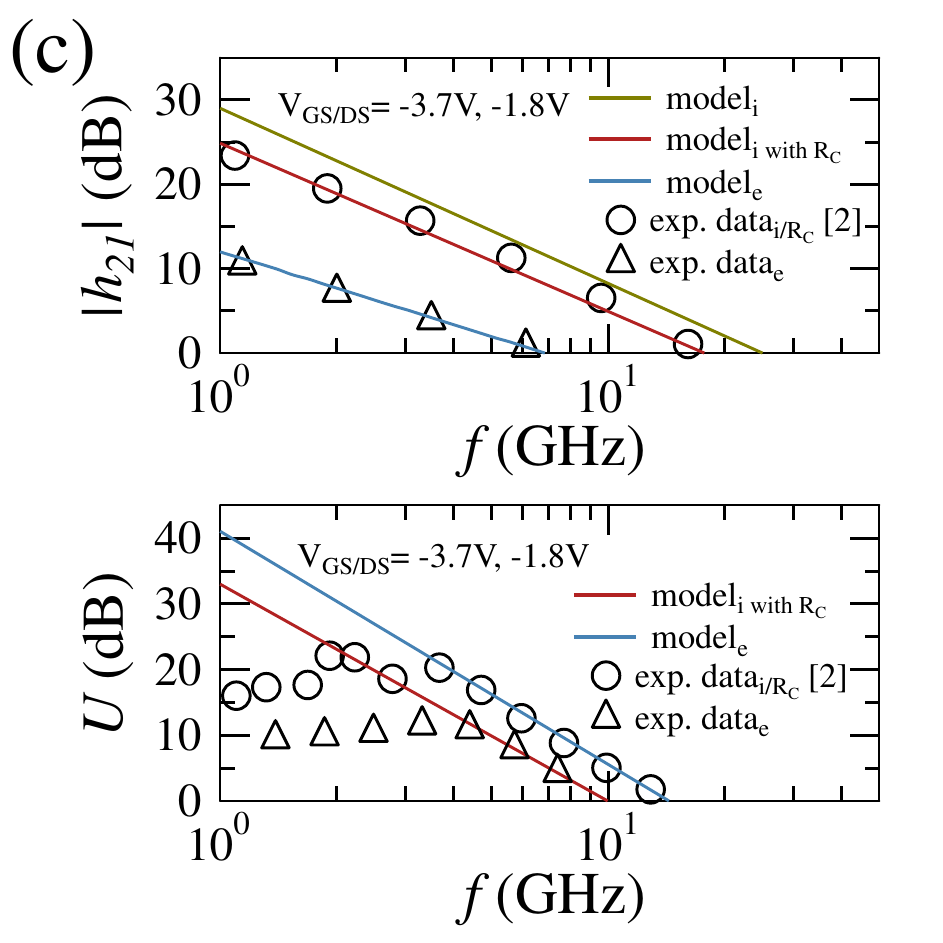}}
	\hspace{5mm}
	\subfloat{\includegraphics[height=0.415\textwidth]{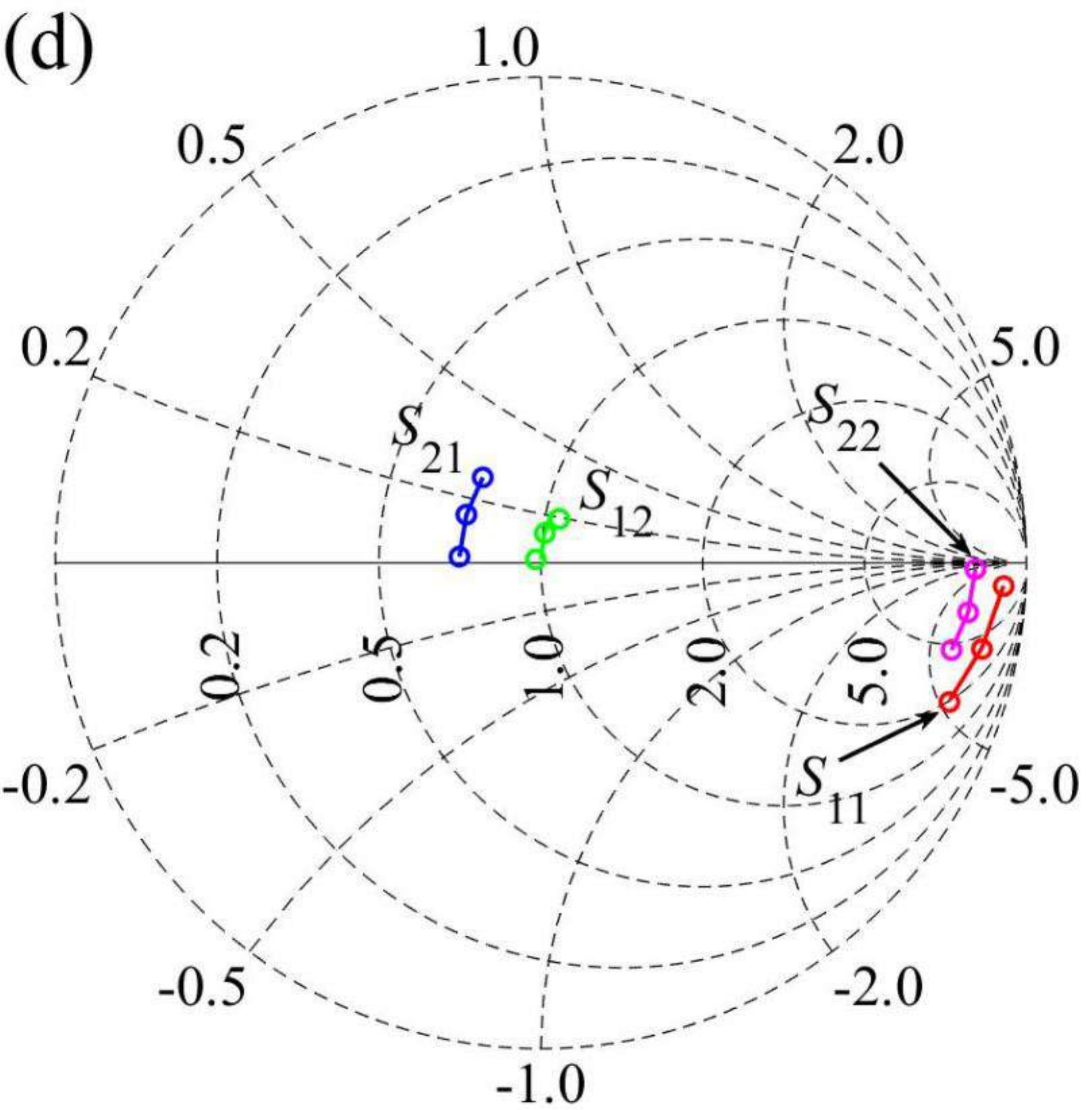}}
	\caption{\small{HF FoM for devices from $T_{\rm 1}$ \cite{ZhuPar16}. \textbf{(a)}, \textbf{(c)} Top (bottom): extrinsic and intrinsic $|h_{\rm 21}|$ ($MAG/MSG$ and $U$) $vs.$ frequency. \textbf{(b)}, \textbf{(d)} $S$-parameters. Data in (a) and (b) correspond to the $T_{\rm 1}$ \SI{250}{\nano\meter}-long device \cite{ZhuPar16}. Data in (c) and (d) correspond to the $T_{\rm 1}$ \SI{500}{\nano\meter}-long device \cite{ZhuPar16}. Markers are experimental data and solid lines correspond to modeling results obtained here.}}
	\label{fig:fT_fMAX_4}
\end{figure}

\vspace{-4mm}
\begin{figure}[!htb] 
	\centering
	\hspace{-0.15cm}
	\subfloat{\includegraphics[height=0.415\textwidth]{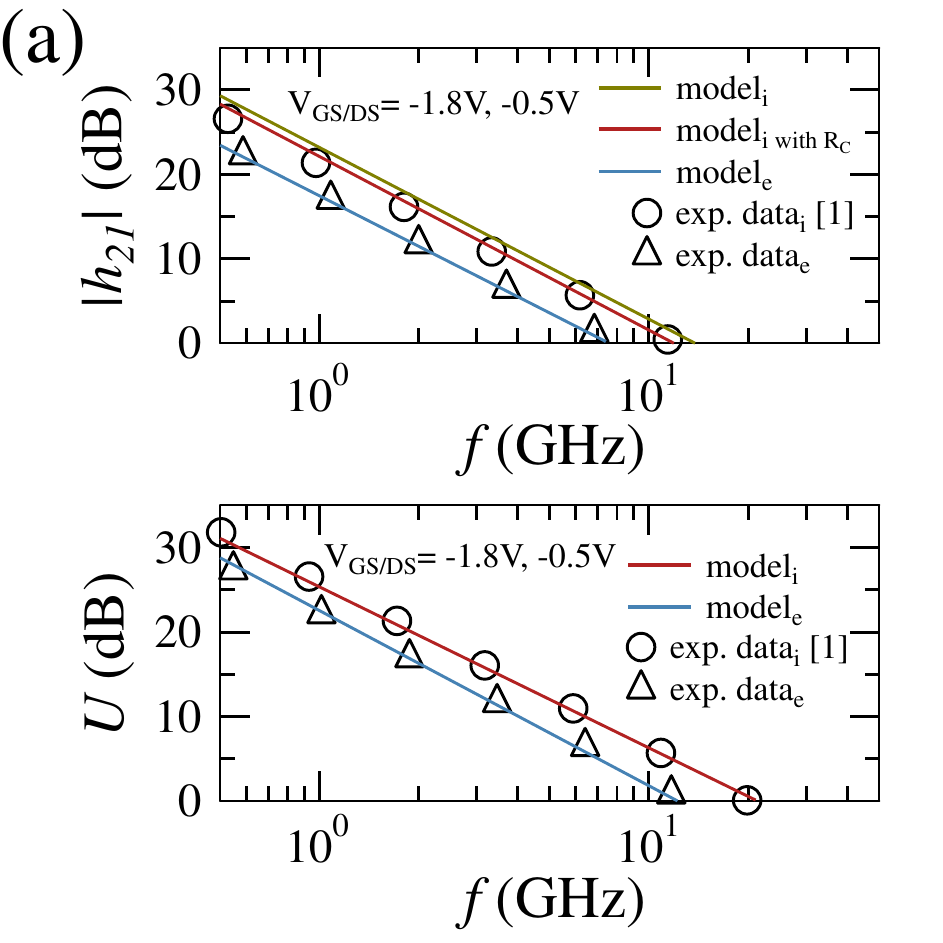}}
	\hspace{5mm}
	\subfloat{\includegraphics[height=0.415\textwidth]{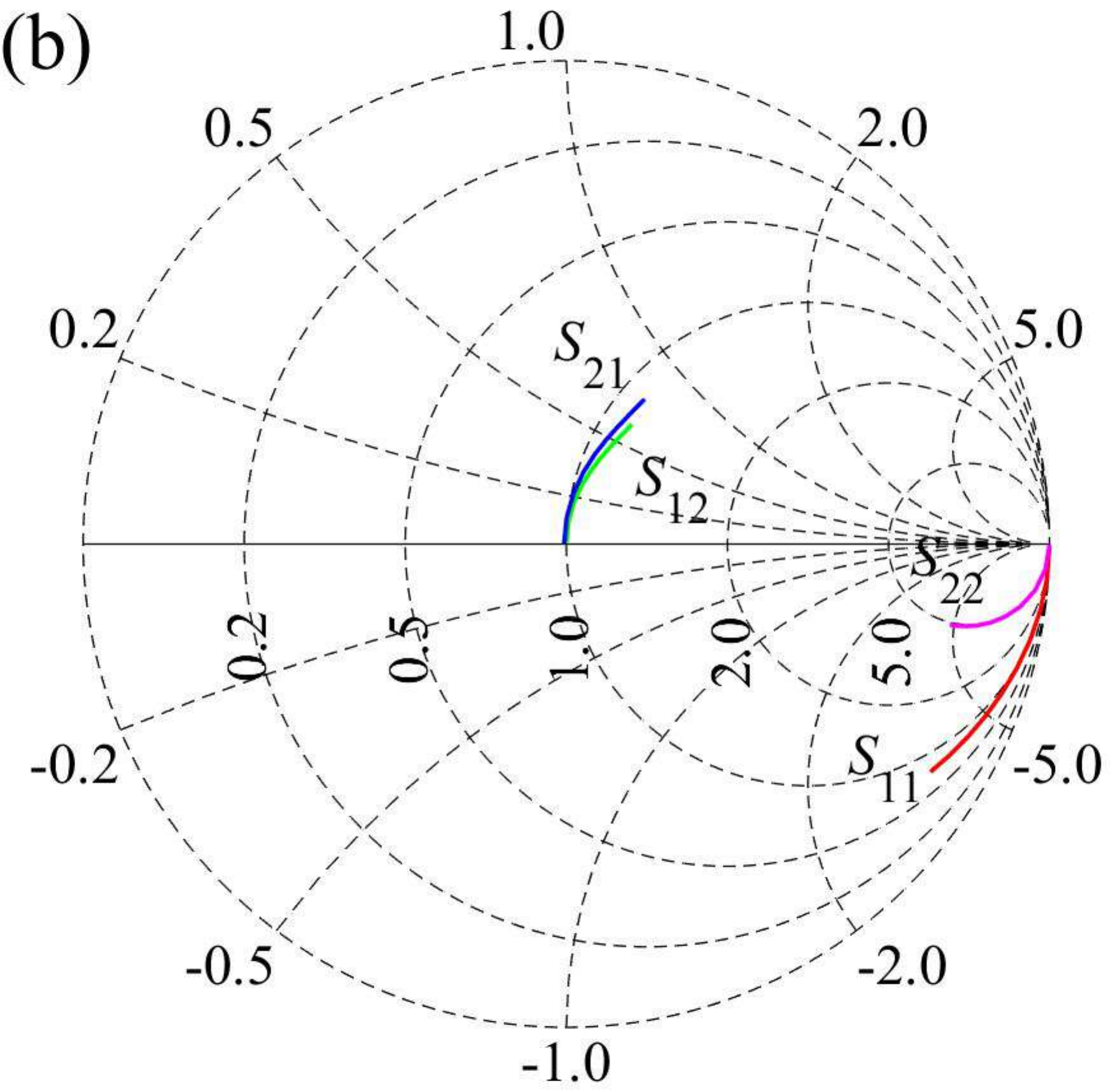}}
	\caption{\small{HF FoM for a \SI{300}{\nano\meter}-long device in \cite{WanWan14}. \textbf{(a)} Top (bottom): extrinsic and intrinsic $|h_{\rm 21}|$ ($U$) $vs.$ frequency. Markers are experimental data reported in \cite{WanWan14} and solid lines correspond to modeling results obtained here, \textbf{(b)} Synthetic $S$-parameters generated with the experimentally calibrated model.}}
	\label{fig:fT_fMAX_3}
\end{figure}

\begin{figure}[!htb] 
	\centering
	\hspace{-0.15cm}
	\subfloat{\includegraphics[height=0.415\textwidth]{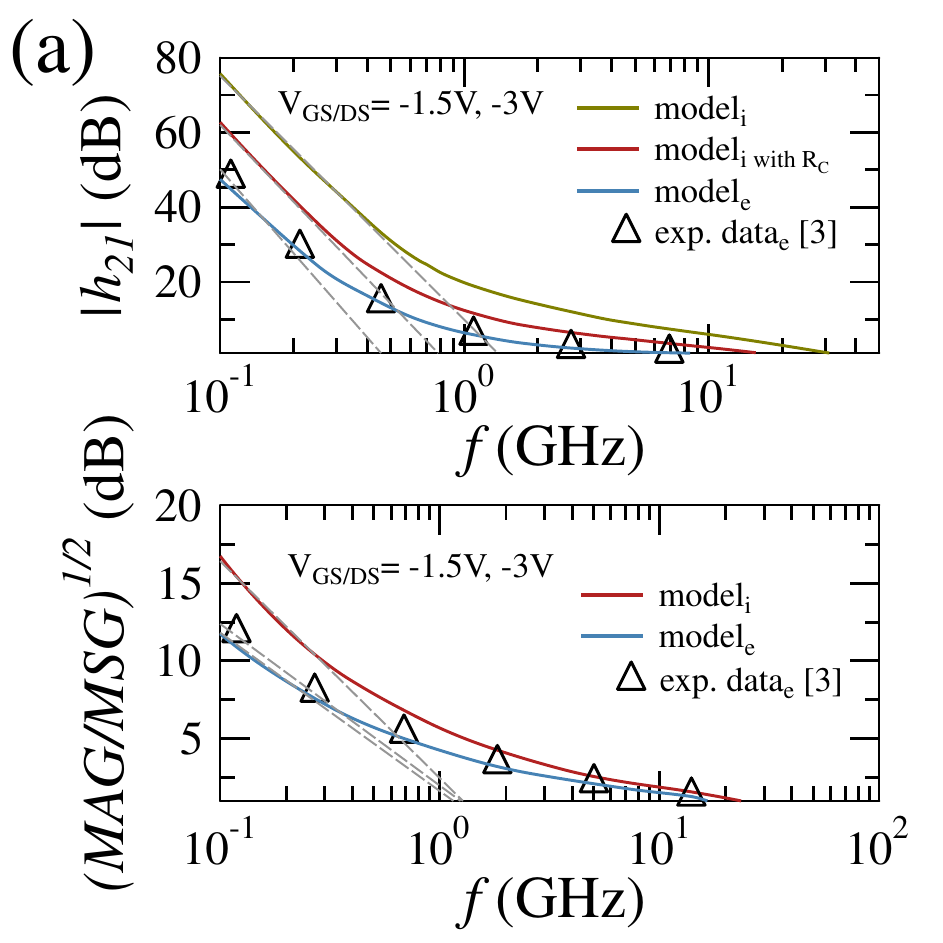}}
	\hspace{5mm}
	\subfloat{\includegraphics[height=0.415\textwidth]{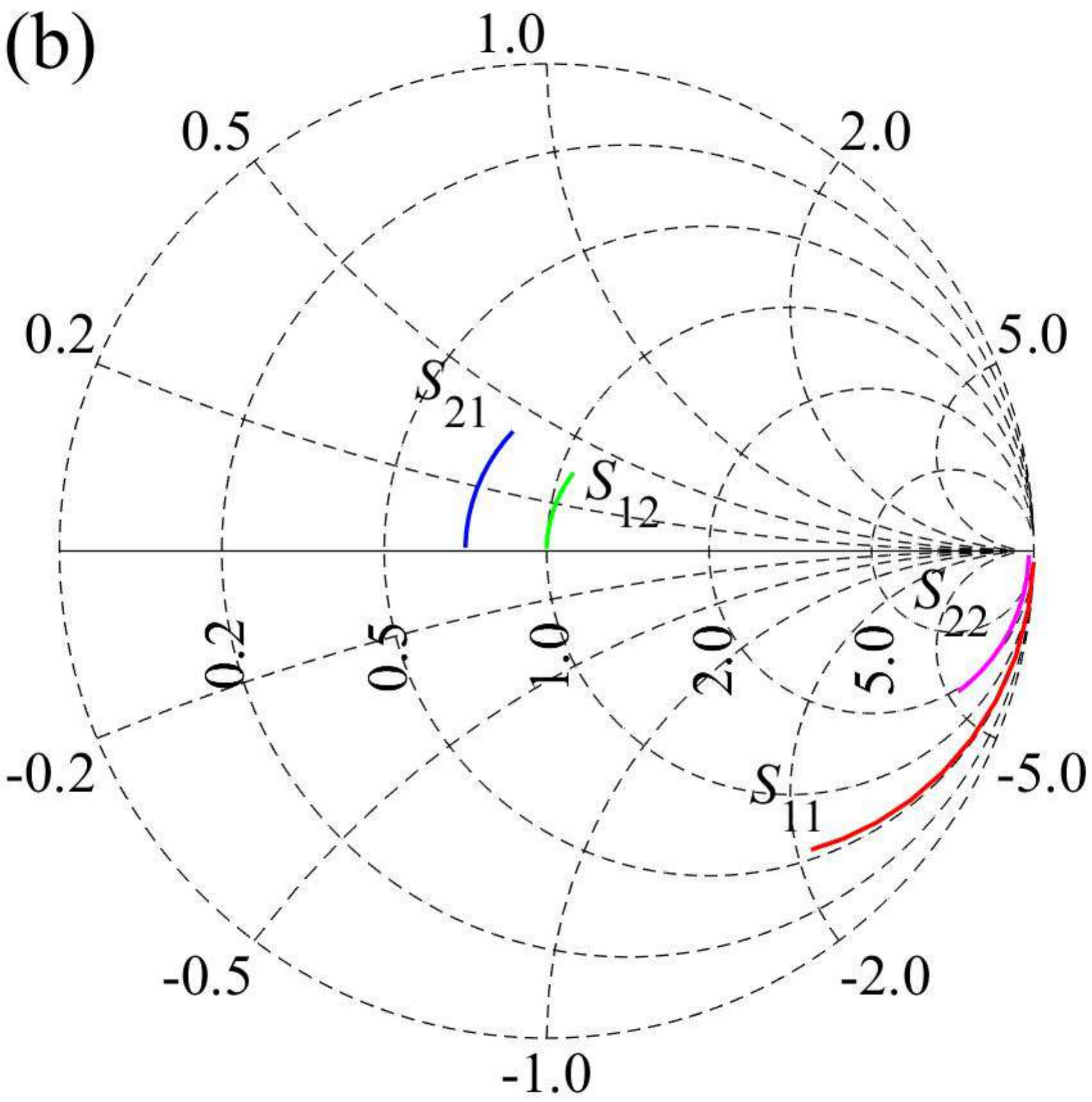}}
	\caption{\small{HF FoM for a \SI{400}{\nano\meter}-long device in \cite{LiTia18}. \textbf{(a)} Top (bottom): extrinsic and intrinsic $|h_{\rm 21}|$ (($MAG/MSG)^{1/2}$) $vs.$ frequency. Markers are experimental data reported in \cite{LiTia18} and solid lines correspond to modeling results obtained here. Dashed lines show the linear extrapolation of the curves. \textbf{(b)} Synthetic $S$-parameters generated with the experimentally calibrated model.}}
	\label{fig:fT_fMAX_5}
\end{figure}

\begin{table} [!htb] 
	\begin{center}	
		\caption{\small{Intrinsic and extrinsic HF FoMs of BPFETs obtained from experimental data and with the proposed small-signal EC.}}
		\scalebox{0.76}{
		\begin{tabular}{c||c|c||c|c||c|c}
			
\makecell{Parameter\\(\SI{}{\giga\hertz})} & \multicolumn{2}{c||}{\makecell{$T_{\rm 1}$ \cite{ZhuPar16}\\\SI{250}{\nano\meter} (\SI{500}{\nano\meter})\\$V_{\rm GS}=\SI{-1.5}{\volt}$ ($\SI{-3.7}{\volt})$,\\$V_{\rm DS}=\SI{-0.5}{\volt}$ ($\SI{-1.8}{\volt})$}} & \multicolumn{2}{c||}{\makecell{$T_{\rm 2}$ \cite{WanWan14}\\\SI{300}{\nano\meter}\\$V_{\rm GS}=\SI{-1.8}{\volt}$,\\$V_{\rm DS}=\SI{-0.5}{\volt}$}} & \multicolumn{2}{c}{\makecell{$T_{\rm 3}$ \cite{LiTia18}\\\SI{400}{\nano\meter}\\$V_{\rm GS}=\SI{-1.5}{\volt}$,\\$V_{\rm DS}=\SI{-3}{\volt}$}} \\ \hline \hline
			
& \cite{ZhuPar16} & This work & \cite{WanWan14} & This work & \cite{LiTia18} & This work  \\ \hline 
			
$f_{\rm{T,i}}$/$f_{\rm{MAX,i}}$ & \SI{15.8}{}/\SI{17.6}{} & \SI{15.3}{}/\SI{19}{} & \SI{12.1}{}/\SI{20}{} & \SI{11.9}{}/\SI{21.1}{} & -/- & \SI{15.6}{}/\SI{23.5}{}   \\  

 & (\SI{17.7}{}/\SI{14.4}{}) & (\SI{17.5}{}/\SI{14.5}{}) & & & &  \\

$f_{\rm{T,e}}$/$f_{\rm{MAX,e}}$ & \SI{5.9}{}/\SI{10.7}{} & \SI{5.7}{}/\SI{10.8}{} & \SI{8}{}/\SI{12}{} & \SI{7.4}{}/\SI{11.8}{} & \SI{8}{}/\SI{17}{} & \SI{8.4}{}/\SI{16.5}{} \\ 

 & (\SI{6.9}{}/\SI{10.1}{}) & (\SI{7}{}/\SI{10.3}{}) & & & &  \\ 
		\end{tabular} \label{tab:FoMs}
	}
	\end{center}
\end{table}

\vspace{-3mm}
Figs. \ref{fig:fT_fMAX_4}-\ref{fig:fT_fMAX_5} present the results obtained from the small-signal EC of Fig. \ref{fig:ckt} and the experimental data at the range of frequencies reported in \cite{WanWan14}-\cite{LiTia18} for different BPFET technologies. Subscripts correspond to intrinsic ($i$) and extrinsic ($e$) data. The latter does not include pad-parasitics. The model describes well the HF short-circuit current gain $h_{\rm{21}}$, the unilateral power gain $U$ as well as the ratio of the maximum available gain (MAG) to the maximum stable gain (MSG), i.e., the power gain, of the fabricated devices \cite{WanWan14}-\cite{LiTia18}. Table \ref{tab:FoMs} reports the intrinsic and extrinsic HF FoMs obtained with experimental and simulation data where a maximum relative error of $\SI{7.9}{}\%$ has been achieved for the $f_{\rm{MAX,i}}$ of the $T_{\rm 1}$ \SI{250}{\nano\meter}-long device \cite{ZhuPar16}. In contrast to \cite{WanWan14} and \cite{LiTia18}, in this work	 $f_{\rm{T}}$ and $f_{\rm{MAX}}$ have been reported from the linear extrapolation of the curve at $\SI{20}{\decibel/dec}$ towards $\SI{0}{\decibel}$ of the corresponding gain curve. Furthermore, Fig. \ref{fig:fT_fMAX_4}(d) validates the proposed model with experimental S-parameters. Hence, the EC indicated in Fig. \ref{fig:ckt} is an efficient option for modeling the HF performance of BPFETs at specific bias points and different frequencies. The modeling approach used here enables to evaluate the impact of metal-channel interface effects on the intrinsic devices performance by obtaining $f_{\rm{T,i}}$ without including $R_{\rm{dc/sc}}$ in the EC as shown in Figs. \ref{fig:fT_fMAX_4}(a), \ref{fig:fT_fMAX_4}(c), \ref{fig:fT_fMAX_3}(a) and \ref{fig:fT_fMAX_5}(a): the larger the $R_{\rm{dc/sc}}$ the more degraded the intrinsic HF performance is. S-parameters obtained with the EC in Fig. \ref{fig:ckt} of the studied devices \cite{WanWan14}-\cite{LiTia18} are shown in Figs. \ref{fig:fT_fMAX_4}(b), \ref{fig:fT_fMAX_4}(d), \ref{fig:fT_fMAX_3}(b) and \ref{fig:fT_fMAX_5}(b).

\begin{figure}[!htb] 
	\centering
	\hspace{-0.15cm}
	\subfloat{\includegraphics[height=0.41\textwidth]{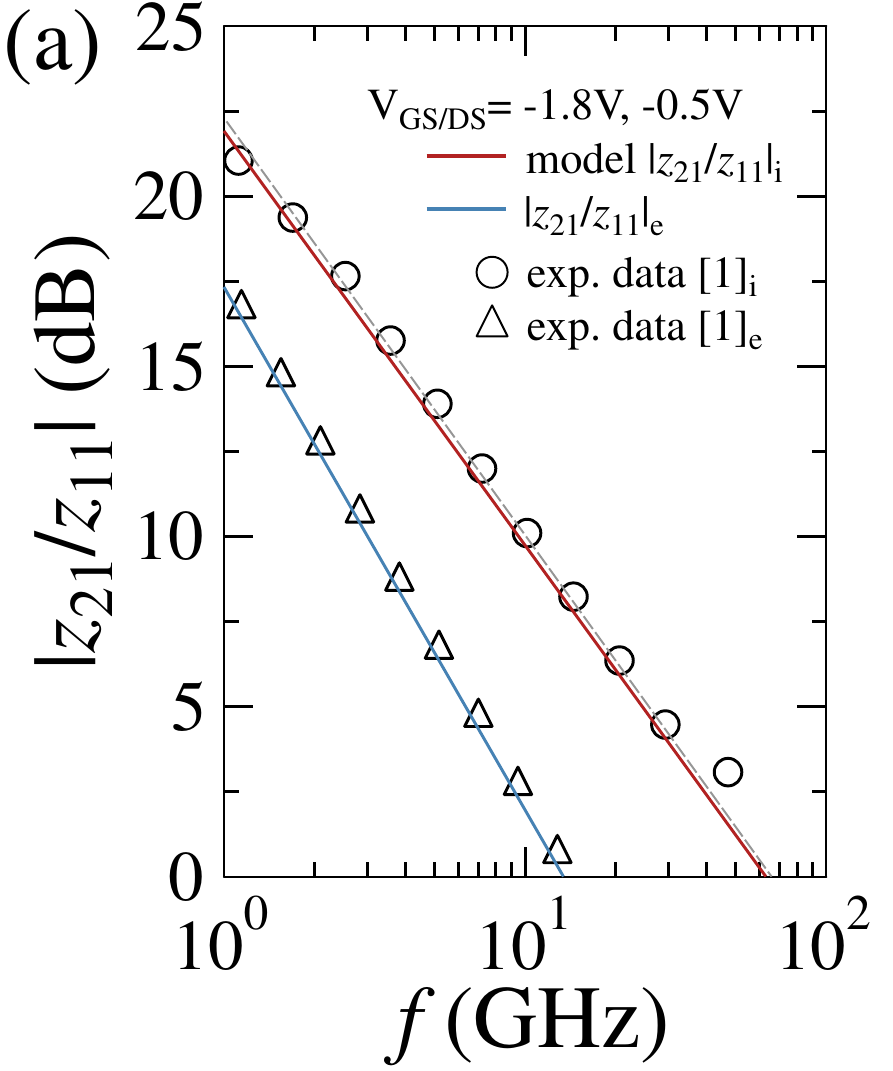}}
	\hspace{7mm}
	\subfloat{\includegraphics[height=0.41\textwidth]{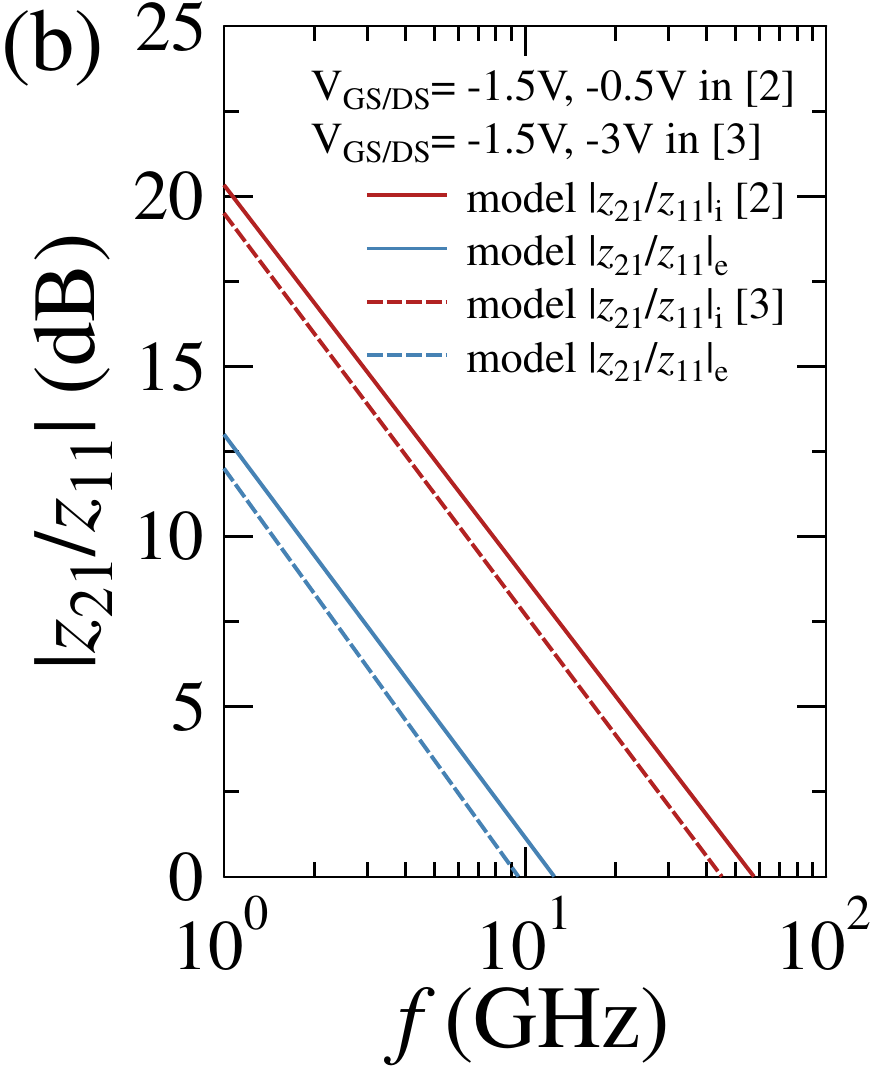}}
	\caption{\small{Experimental (markers) and simulation (lines) results of the intrinsic and extrinsic open-circuit voltage gain $vs.$ frequency of the \textbf{(a)} $T_{\rm 2}$ device \cite{WanWan14} and \textbf{(b)} of the $T_{\rm 1}$ \cite{ZhuPar16} and $T_{\rm 3}$ \cite{LiTia18} BPFETs.}}
	\label{fig:Z21_Z11}
\end{figure}

As shown in Fig. \ref{fig:Z21_Z11}, the intrinsic experimental (modeling) response of the open-circuit voltage gain $|z_{\rm 21}/z_{\rm 11}|$ of the $T_{\rm 2}$ device \cite{WanWan14} is $\SI{65.8}{\giga\hertz}$ ($\SI{63}{\giga\hertz}$) and the extrinsic one is $\SI{14.2}{\giga\hertz}$ ($\SI{13.4}{\giga\hertz}$). Hence, these results, along with the previous ones (cf. Figs. 2-4), indicate that the model is efficient in describing BPFETs at HF short- and open-circuit conditions. Synthetic data of the $\SI{250}{\nano\meter}$-long \cite{ZhuPar16} and  $\SI{500}{\nano\meter}$-long devices \cite{LiTia18} shown in Fig. \ref{fig:Z21_Z11}(b) indicate a high extrinsic voltage gain ($>$ $\SI{5}{\decibel}$) up to $\sim$ $\SI{5}{\giga\hertz}$ for the shortest device \cite{ZhuPar16} and up to $\sim$ $\SI{3.5}{\giga\hertz}$ for the longest BPFET \cite{LiTia18}.

\section{High-frequency circuit design}
\label{ch:s_III}

\begin{figure}[!htb]
	\centering
	\includegraphics[width=0.595\textwidth]{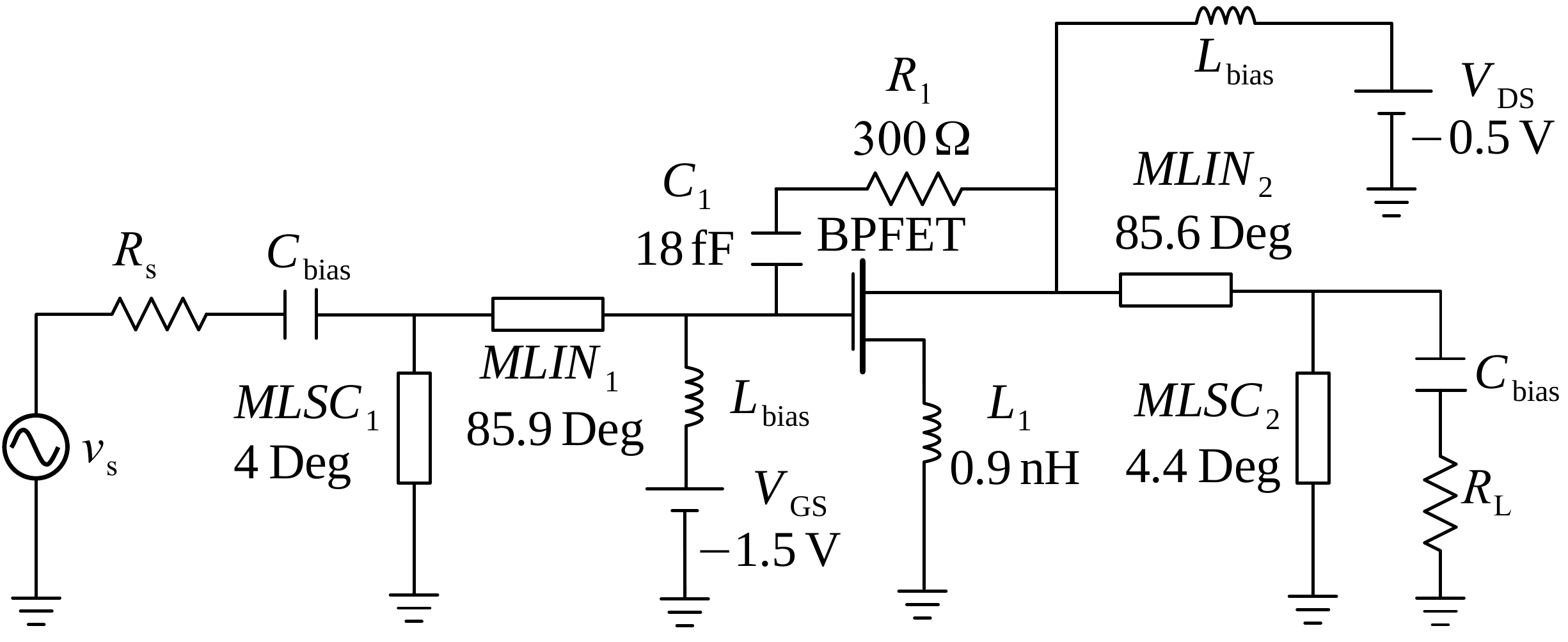}
	\caption{\small{Schematic representation of the single-stage maximum gain amplifier based on BPFET in common-source topology.}}
	\label{fig:MGA}
\end{figure}

The small-signal model proposed in this work enables the design of HF circuits, e.g., multipliers, amplifiers, phase-shifters. The schematic of a single-stage maximum gain amplifier based on the $T_{\rm 1}$ $\SI{250}{\nano\meter}$-long device \cite{ZhuPar16} and designed by using the EC suggested in this work in common-source configuration and with a feedback topology is presented in Fig. \ref{fig:MGA}. The amplifier has been designed to operate at a frequency equal to $\SI{2.4}{\giga\hertz}$ and at the same bias point in which the model has been validated for this device \cite{ZhuPar16} (cf. Section \ref{ch:s_II}). QS conditions prevail in the device at this frequency since it is lower than the corresponding $f_{\rm{T,i}}$ for this technology \cite{ZhuPar16} (see Table \ref{tab:FoMs}), as predicted elsewhere \cite{DasMah20}. In contrast to \cite{ChoYog16} were a BPFET-based HF amplifier has been designed by coupling its output at $\SI{1}{\mega\ohm}$, in this work the device output has been matched to the $\SI{50}{\ohm}$-standard for RF environments.

\begin{figure}[!htb]
	\centering
	\hspace{-0.15cm}
	\subfloat{\includegraphics[height=0.41\textwidth]{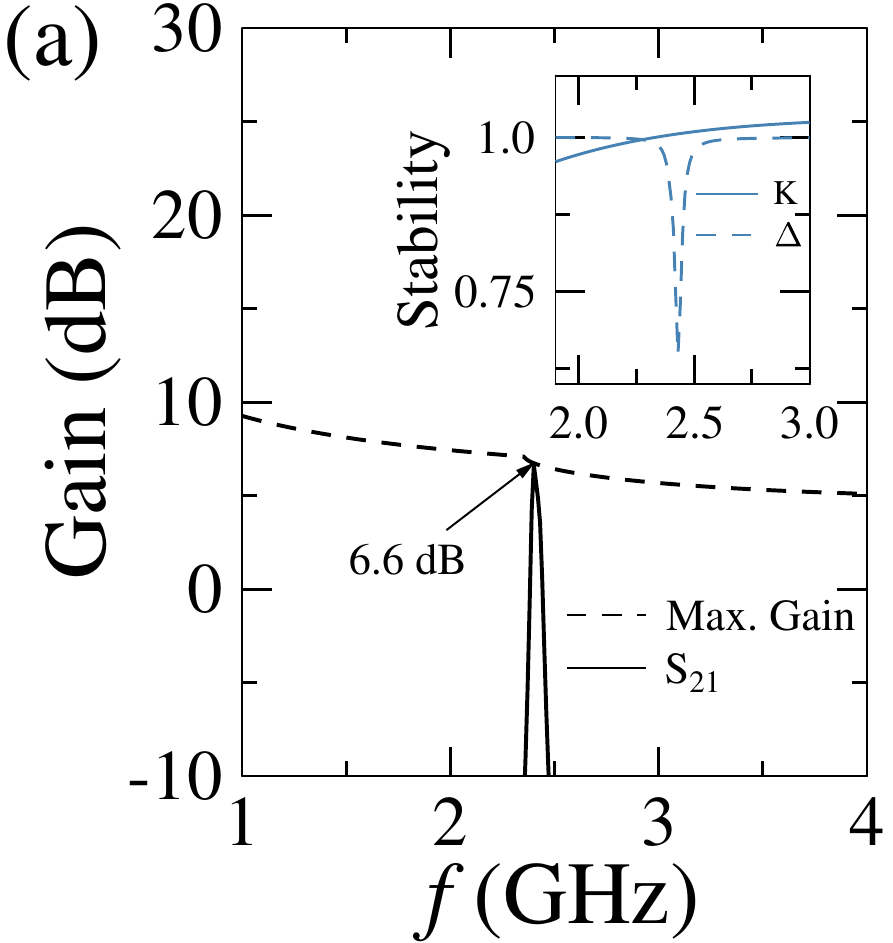}}
	\hspace{4mm}
	\subfloat{\includegraphics[height=0.41\textwidth]{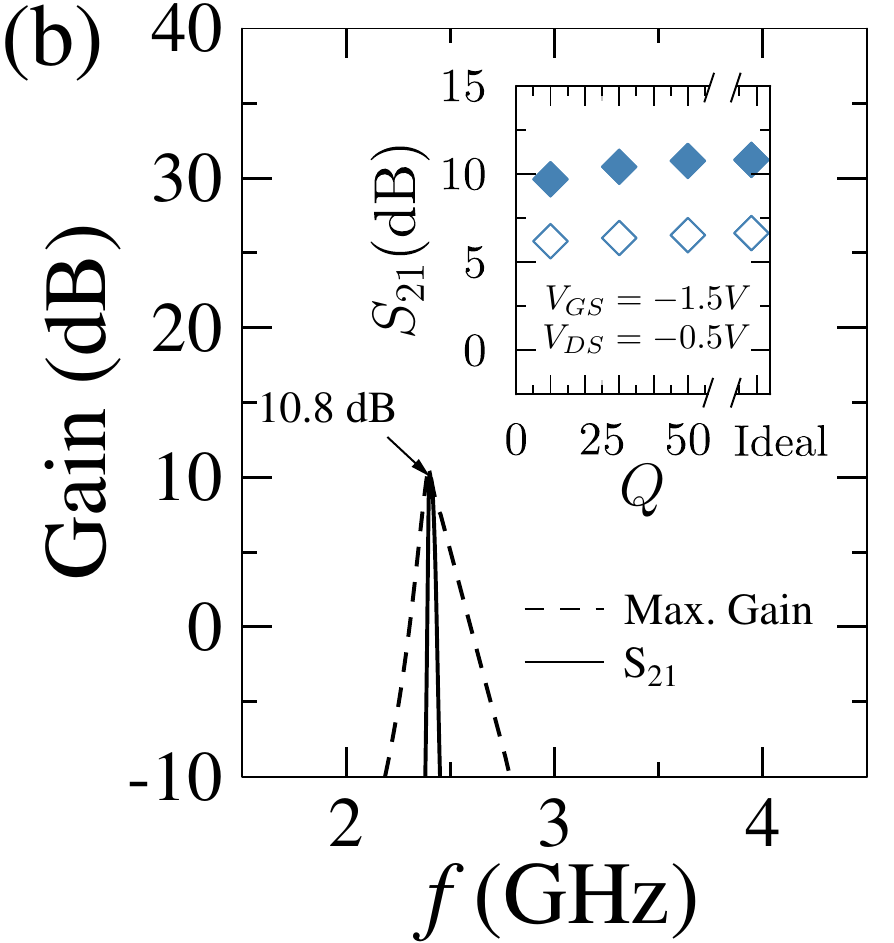}}
	\caption{\small{Available Gain for BPFET-based amplifiers: \textbf{(a)} single-stage and \textbf{(b)} double-stage. Inset in \textbf{(a)} shows stability coefficients where the \textit{x}-axis corresponds to $f$ in \SI{}{\giga\hertz}. Inset in \textbf{(b)} shows $S_{\rm 21}$ as a function of $Q$-factor for $L_{\rm{1}}$ (cf. Fig. \ref{fig:MGA}) of the single-stage (empty markers) and double-stage (filled markers) designs at $\SI{2.4}{\giga\hertz}$.}}
	\label{fig:MGA_Graphic}
\end{figure}

The stability conditions \cite{PozD11} for the device ($K>\SI{1}{}$ and $\Delta<\SI{1}{}$ at the frequency range between $\SI{2.35}{\giga\hertz}$ and $\SI{2.6}{\giga\hertz}$), as shown in the inset of Fig. \ref{fig:MGA_Graphic}(a) have been obtained using a drain-gate shunt $RC$ feedback network and also with a source series high-Q inductor $L$. The maximum width obtained for the feedback network components is $\sim$ $\SI{5}{\micro\meter}$ which enables monolithic integration along with the active device since the reference substrate thickness is much larger ($\sim$ $\SI{125}{\micro\meter}$) \cite{ZhuPar16}. The corresponding plots, inset of Fig. \ref{fig:MGA_Graphic}(a), indicate that the device is stable in the range of frequency between $\SI{2.35}{\giga\hertz}$ and $\SI{2.6}{\giga\hertz}$. Microstrip lines ($MLIN_{\rm 1}$ and $MLIN_{\rm 2}$) and short circuit stubs ($MLSC_{\rm 1}$ and $MLSC_{\rm 2}$) have been used for the matching networks at the input and output, designed so that $S_{\rm 11}$ and $S_{\rm 22}$ are lower than $-\SI{10}{\decibel}$ at $\SI{2.4}{\giga\hertz}$.

Fig. \ref{fig:MGA_Graphic} shows maximum available gain ($G_{\rm{A}}$) and $S_{\rm 21}$ curves versus of single- and two-stage BPFET-based RF amplifiers. The latter has been obtained by connecting two single-stage amplifiers (cf. Fig. \ref{fig:MGA}) in cascade. The correct design of feedback, stability and matching networks enable to improve the poor $S_{\rm{21}}$ device response (see Fig. \ref{fig:fT_fMAX_4}(b)) to a selective positive response of the single- (double-) stage amplifier at $\SI{2.4}{\giga\hertz}$ with a bandwidth of \SI{12}{\mega\hertz} (\SI{14}{\mega\hertz}) and a $S_{\rm 21}$ of $\SI{6.6}{\decibel}$ ($\SI{10.8}{\decibel}$). A similar narrow bandwidth has been achieved for a different BPFET-based RF amplifier designed elsewhere \cite{ChoYog16}. The closeness between $S_{\rm 21}$ and the maximum gain of the single-stage design is $<\SI{0.5}{}\%$ at $\SI{2.4}{\giga\hertz}$ and the frequency range in which the double-stage amplifier is above $\SI{3}{\decibel}$ is between $\SI{2.392}{\giga\hertz}$ and $\SI{2.408}{\giga\hertz}$. The results suggest that BPFET-based amplifiers can be used, e. g., in RF front-ends in order to share the frequency spectrum efficiently and to obtain a dynamic access to the spectrum at narrow bandwidths \cite{BeaLau20}. The gain achieved at $\SI{2.4}{\giga\hertz}$ with the single-stage amplifier design indicates that BPFET-based HF amplifiers can compete with other more studied 2D technologies at similar operation frequencies, e.g., fabricated graphene FET-based single-stage circuits \cite{YehC14}, \cite{HanT17} yield only $\sim$$\SI{1.5}{\decibel}$ more, at $\SI{2.4}{\giga\hertz}$ and  $\SI{2.5}{\giga\hertz}$, respectively, than the results shown here. Inset of Fig. \ref{fig:MGA_Graphic}(b) shows a study of the impact of the quality factor ($Q$) of the inductors $L_{\rm{1}}$ on the amplifiers performance by using values of $Q=$ \SI{10}{}, \SI{30}{}, \SI{50}{}, in addition to the ideal $Q$-case previously considered. The gain performance of the single-stage amplifier is very close to the ideal case, obtaining \SI{6}{\decibel} in the worst case ($Q$ = \SI{10}{}). For the two-stage amplifier the impact of $Q$ diminishes the gain \SI{1.1}{\decibel} with respect to the ideal case.

\vspace{-2mm}
\section{Conclusion}

The small-signal EC presented in this work is an efficient tool  for modeling AC performances of black-phosphorous field-effect transistors. This model has described properly the intrinsic and extrinsic HF figures of merit of devices from different technologies over the range of frequencies in which these BPFETs are suitable for HF current and power amplification. Furthermore, this model has predicted with a difference $\SI{<5}{\percent}$ the reported values of the intrinsic and extrinsic open-circuit voltage gain curves of a fabricated device. The small-signal model proposed here offers the first successful description of the experimental HF performance of BPFET technologies. Single- and two-stage BPFET-based amplifiers designed with an experimentally-calibrated model have gains equal to $\SI{6.6}{\decibel}$ and $\SI{10.8}{\decibel}$, respectively, at 2.4 GHz, with narrow bandwidths each of them. Hence, BPFET-based gain amplifiers are good candidates to develop highly selective RF front-end designs.

\end{document}